\documentclass{sig-alternate-2013}
\usepackage{etoolbox}
\makeatletter
\patchcmd{\maketitle}{\@copyrightspace}{}{}{}
\makeatother

\newtheorem{observation}{Observation}

\newtheorem{theorem}{Theorem}[section]

\newtheorem{lemma}[theorem]{Lemma}

\newtheorem{definition}[theorem]{Definition}
\newtheorem{example}[theorem]{Example}

\usepackage{algorithm}
\usepackage{comment}

\usepackage{algorithmic}
\usepackage{xcolor}

\newcommand{\G}{\ensuremath{\mathcal{G}}}

\newcommand{\DC}{\ensuremath{\textsc{DC}}}
\newcommand{\CC}{\ensuremath{\textsc{CC}}}

\newcommand{\BC}{\ensuremath{\textsc{BC}}}
\newcommand{\GED}{\ensuremath{\textsc{GED}}}

\newfont{\mycrnotice}{ptmr8t at 7pt}
\newfont{\myconfname}{ptmri8t at 7pt}
\let\crnotice\mycrnotice%
\let\confname\myconfname%

\permission{Permission to make digital or hard copies of all or part of this work for personal or classroom use is granted without fee provided that copies are not made or distributed for profit or commercial advantage and that copies bear this notice and the full citation on the first page. Copyrights for components of this work owned by others than the author(s) must be honored. Abstracting with credit is permitted. To copy otherwise, or republish, to post on servers or to redistribute to lists, requires prior specific permission and/or a fee. Request permissions from permissions@acm.org.}
\conferenceinfo{FOMC'14,}{August 11, 2014, Philadelphia, PA, USA. \\
{\mycrnotice{Copyright is held by the owner/author(s). Publication rights licensed to ACM.}}}
\copyrightetc{ACM \the\acmcopyr}
\crdata{978-1-4503-2984-2/14/08\ ...\$15.00.\\
http://dx.doi.org/10.1145/2634274.2634277
}

\begin{document}

\title{Understanding Dynamic Graphs:\\The Case for Centrality Distance}

\title{Modeling and Measuring Graph Similarity:\\The Case for Centrality Distance}

\author{Matthieu Roy$^{1,2}$, Stefan Schmid$^{1,3,4}$, Gilles Tredan$^{1,2}$\\
\small $^1$ CNRS, LAAS, 7 avenue du colonel Roche, F-31400 Toulouse, France \\
\small $^2$ Univ de Toulouse, LAAS, F-31400 Toulouse, France\\
\small $^3$ Univ de Toulouse, INP, LAAS, F-31400 Toulouse, France\\
\small $^4$ TU Berlin \& T-Labs, Berlin, Germany}

\date{}
\maketitle

\sloppy

\begin{abstract}
The study of the topological structure of complex networks
has fascinated researchers for several decades, and today we have a
fairly good understanding of the types and reoccurring characteristics of many different
complex networks. However, surprisingly little is known today about
models to compare complex graphs, and quantitatively measure their similarity.

This paper proposes a natural similarity measure for complex networks:
\emph{centrality distance}, the difference  between two graphs with respect to a given node centrality.
Centrality distances allow to take into account the specific roles of the different nodes in the network,
and have many interesting applications. As a case study,
we consider the closeness centrality in more detail,
and show that closeness centrality distance can be used to effectively
distinguish between randomly generated and 
actual evolutionary
paths of two dynamic social networks. 
\end{abstract}




%\gilles{commentaire} \matthieu{commentaire} \stefan{commentaire}~\\

\section{Introduction}
\label{sec:introduction}

How similar are two graphs $G_1$ and $G_2$? Surprisingly, today,
we do not have good measures to answer this question. In graph theory,
the canonical measure to compare two graphs is the \emph{Graph Edit Distance ($\GED$)}~\cite{ged-survey}:
informally, the GED $d_{\GED}(G_1,G_2)$ between two graphs $G_1$ and $G_2$
is defined as the minimal number of \emph{graph edit operations} that are
needed to transform $G_1$ into $G_2$. The specific set of allowed graph edit operations
depends on the context, but typically includes some sort of
node and link insertions and deletions.

While graph edit distance metrics play an important role in computer graphics
and are widely applied to
pattern analysis and recognition, we argue that the graph edit distance is not well-suited
for measuring similarities between natural and complex networks. The set of graphs
at a certain graph edit distance $d$ from a given graph $G$, are very diverse and seemingly
unrelated: the characteristic structure of $G$ is lost.

A good similarity measure can have many important applications. For instance, a similarity
measure can be a fundamental tool for the study of dynamic networks, answering questions like:
\emph{Do these two complex networks have a common ancestor?} Or: \emph{What is a likely successor network for a given network?}
While the topological properties of complex networks have fascinated researchers for many decades,
(e.g., their connectivity~\cite{bar-alb,pref-attach},
their constituting motifs~\cite{motifs}, their clustering~\cite{wattsstrogatz} or community patterns~\cite{community}),
today, we do not have a good understanding of their dynamics over time.

\textbf{Our Contributions.}
This paper initiates the study of graph similarity measures
for complex networks which go beyond simple graph edit distances. In particular, we introduce the notion of
\emph{centrality distance} $d_{C}(G_1,G_2)$, a graph similarity measure
based on a \emph{node centrality} $C$.

We argue that centrality-based distances are attractive similarity measures as they are naturally \emph{node-oriented}.
This stands in contrast to, e.g., classic graph isomorphism based measures which apply only to anonymous graphs;
in the context of dynamic complex networks, nodes typically do represent real objects and are not anonymous!

We observe that the classic graph edit distance can be seen as a special case of centrality distance: the graph edit distance is equivalent to the
centrality distance where $C$ is simply the degree centrality, henceforth referred to as the \emph{degree distance} $d_{\DC}$.
We then discuss alternative centrality distances and, as a case study, explore the \emph{closeness distance} $d_{\CC}$
(based on closeness centrality) in more detail.

In particular, we show that closeness distance has interesting applications in the domain of dynamic network prediction.
As a proof-of-concept, we consider two dynamic social networks: (1) An evolving network representing the human mobility during
a cocktail party, and (2) a Facebook-like Online Social Network (OSN) evolving over time.
%We find that closeness centrality distances can provide interesting insights into the network dynamics.
We show that actual evolutionary paths are far from being random
from the perspective of closeness centrality distance, in the sense that
the distance variation along evolutionary paths is low. 
This can be exploited to distinguish between fake and actual evolutionary
paths with high probability. 

\textbf{Examples.}
To motivate the need for graph similarity measures, let us consider two simple examples.

\begin{example}[Local/Global Scenario]
We consider three graphs $G_1$, $G_2$, $G_3$ over five nodes $\{v_1,v_2,\ldots,v_5\}$: $G_1$ is a line,
where $v_i$ and $v_{i+1}$ are connected in a modulo manner; $G_2$ is a cycle, i.e.,  $G_1$  with
an additional link $\{v_1,v_5\}$; and $G_3$ is  $G_1$ with an additional link $\{2,4\}$.
\end{example}

In this example, we first observe that $G_2$ and $G_3$ have the same graph edit distance to $G_1$: $d_{\GED}(G_1,G_2)=d_{\GED}(G_1,G_3)=1$,
as they contain one additional edge. However, in a social network context, one would intuitively expect $G_3$
to be closer to $G_1$ than $G_2$. For example, in a friendship network a short-range ``\emph{triadic closure}''~\cite{triadic} link may be more likely
to emerge than a long-range link: friends of friends may be more likely to become friends themselves in the future.
Moreover, more local changes are also
expected in mobile environments (e.g., under bounded human mobility and speed).
As we will see, the centrality distance concept introduced in this paper can capture such differences.

%\begin{wrapfigure}{l}{0.6\textwidth}
%%\begin{figure}[h]
%\begin{center}
%\vspace{-.7cm}
%\includegraphics[width=0.5\textwidth]{three-graphs}
%	  \caption{Local/Global Scenario: $G_1$ describes a line graph, $G_2$ describes a shell graph.}\vspace{-.6cm}%
%	\label{fig:stat-ex}
%	\end{center}
%\end{figure}
%\end{wrapfigure}

\begin{figure*}[t]
\centering
\includegraphics[width=0.45\textwidth]{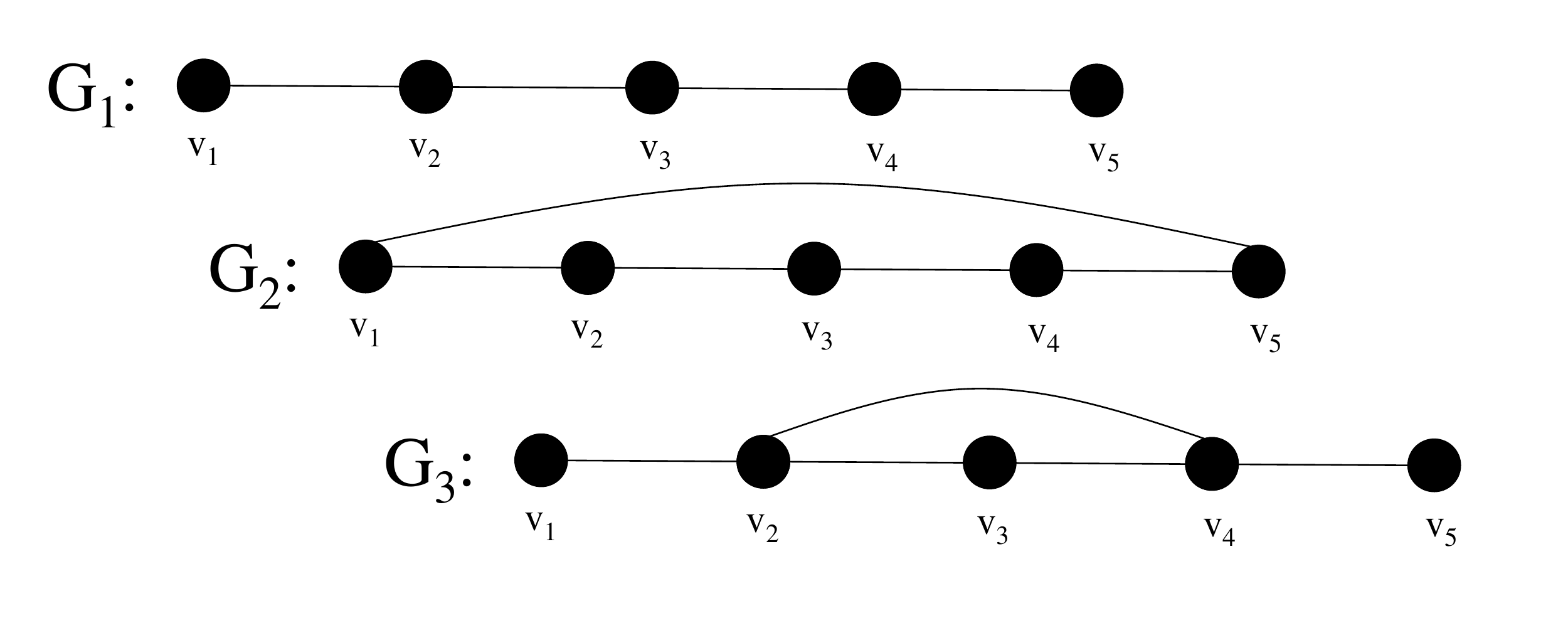}~~~\includegraphics[width=0.45\textwidth]{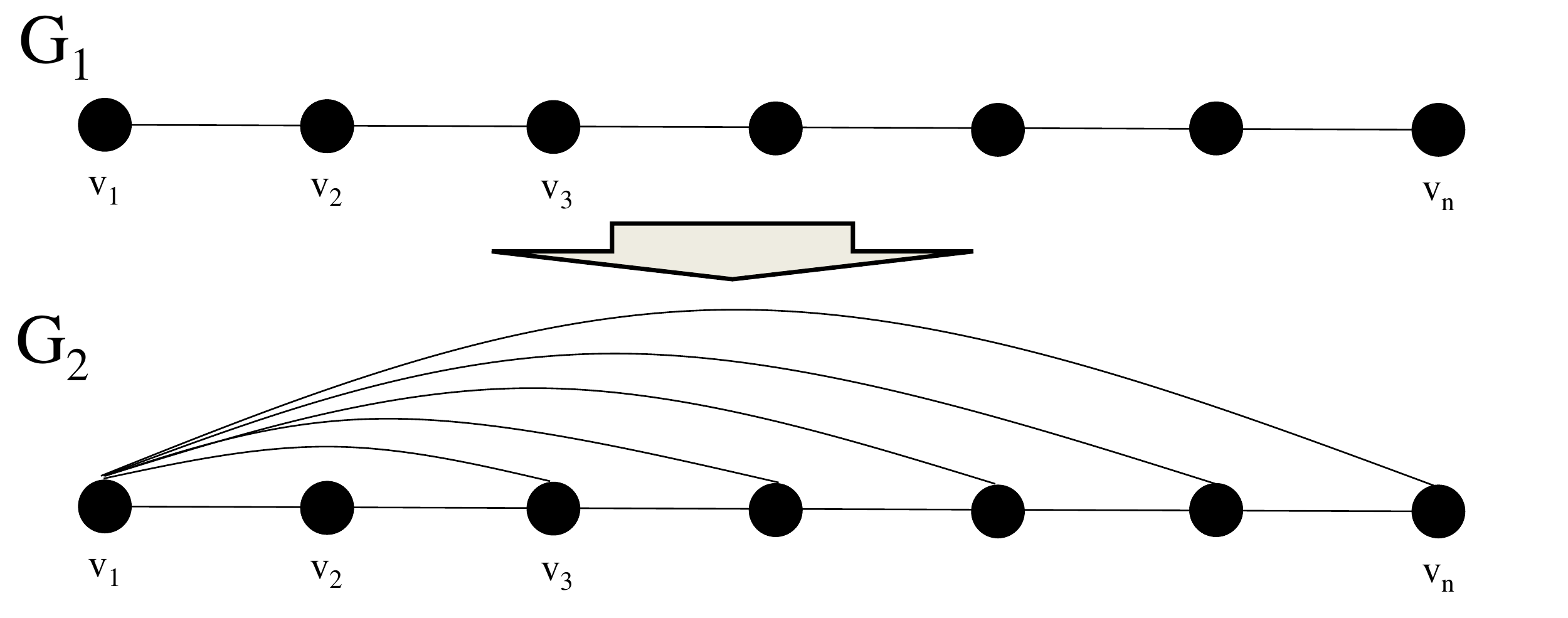}\\
\caption{\emph{Left:} Local/Global Scenario: $G_1$ is a line graph, $G_2$ and $G_3$ are obtained from $G_1$ by adding a link. \emph{Right:}
Evolution Scenario: $G_1$ describes a line graph, $G_2$ describes a shell graph.}
\label{fig:dyn-ex} 	\label{fig:example}
\end{figure*}

\begin{example}[Evolution Scenario]
As a second artificial and very simple example, in this paper we will consider two graphs $G_1$ and $G_2$,
where $G_1$ is a line topology and $G_2$ is a ``shell network'', shown in Figure~\ref{fig:example}.
We ask the question: what is the most likely evolutionary path that would lead from the $G_1$ topology to $G_2$?
\end{example}

Note that the graph edit distance does not provide us with any information about the likely \emph{evolutionary paths} from
$G_1$ to $G_2$, i.e., on the order of the edge insertions: there are many possible orders in which the missing links can be added to $G_1$,
and these orders do not differ in any way.
In reality, however, we often have some expectations on how a graph may have evolved between the two given snapshots $G_1$ and $G_2$.
For example, applying the triadic closure principle to our example, we would expect that the missing links are introduced one-by-one, from left to right.
A similar evolution may also be predicted by a temporal preferential attachment model~\cite{pref-attach}: the degree of the most highly connected
node is likely to grow further in the future.
%\begin{wrapfigure}{r}{0.6\textwidth}
%%\begin{figure}[h]
%\begin{center}
%\vspace{-.7cm}
%\includegraphics[width=0.5\textwidth]{example}
%	  \caption{Evolution Scenario: $G_1$ describes a line graph, $G_2$ describes a shell graph.}\vspace{-.6cm}
%	\label{fig:dyn-ex}
%	\end{center}
%%\end{figure}
%\end{wrapfigure}

The situation may look different in a technological, man-planned network.
For example, adding links from left to right only slowly improves the ``routing efficiency'' of the network:
after the addition of $t$ edges from left to right, the longest
shortest path is $n-t$ hops, for $t<n-1$. A ``better'' evolution of the network can be obtained by adding links to the middle of the network,
reducing the much faster in the beginning: after $t$ edge insertions, the distance is roughly reduced by a factor $t$.

%\textbf{Organization.}
%The remainder of this paper is organized as follows.
%Section~\ref{sec:background} provides the reader with the necessary background
%on centralities. Section~\ref{sec:graph-dist} introduces our centrality distance
%framework. We discuss and compare specific centrality distances in Section~\ref{sec:analysis}.
%Section~\ref{sec:experiments} reports on experimental results with closeness distance in two case studies.
%After reviewing related work in Section~\ref{sec:relwork}, we conclude our
%contribution in Section~\ref{sec:conclusion}.

\section{Model and Background}\label{sec:background}

This paper focuses on \emph{named} (a.k.a.~labeled) graphs $G=(V,E)$: graphs where vertices
$v\in V$ have unique identifiers and are connected via undirected edges $e\in E$.
We focus on \emph{node centralities}, centralities assigning ``importance values''
to nodes $v\in V$.

\begin{definition}[Centrality]
  A \emph{centrality} $C$ is a function $C\colon (G,v) \to \mathbb{R}^{+}$ that
  takes a graph $G=(V,E)$ and a vertex $v\in V(G)$ and returns a positive value
  $C(G,v)$.  The centrality function is defined over all vertices $V(G)$ of a
  given graph $G$. Although we here consider named graphs, we require
    centrality values of vertices to be independent of the vertex's identifier, i.e., centralities
    are unchanged by a permutation of identifiers:
    given any permutation $\pi$ of $V(G)$, $\forall
    v \in V(G), C(\pi(G),\pi(v)) = C(G,v)$, where $\pi(G)=(V(G), \{(\pi(v),\pi(v')) : (v,v')\in E(G)\}$.
\end{definition}

Centralities are a common way to characterize complex networks and their vertices.
Frequently studied centralities include the \emph{degree
centrality} (\DC), the \emph{betweenness centrality} (\BC) and the \emph{closeness centrality}
(\CC), among many more. A node is $\DC$-central if it has many edges: the degree centrality is simply the node degree;
a node is $\BC$-central if it is on many shortest paths: the betweenness centrality is the number of shortest paths going
through the node; and a node is $\CC$-central if it is close to many other nodes: the closeness centrality
measures the inverse of the distances to all other nodes.
Formally:
\begin{enumerate}
\item \emph{Degree Centrality:} For any node $v\in V(G)$ of a network $G$, let $\Gamma(v)$ be the set of neighbors of node $v$:
  $\Gamma(v)=\{ w\in V \textrm{s.t.~}\{v,w\} \in E\}$.
  The \emph{degree centrality} $\DC$ of a
  node $v\in V$ is defined as: $\DC(G,v)=\vert \Gamma(v) \vert$.

\item \emph{Betweenness Centrality:}
   For any pair $(v,w) \in E(G)$, let $\sigma(v,w)$ be the total number of
   different shortest paths between $v$ and $w$, and let $\sigma_x(v,w)$ be the number of shortest paths
  between $v$ and $w$ that pass through $x \in V$.  The betweenness centrality
  $\BC$ of a node $v\in V$ is defined as: $\BC(G,v)=\sum_{x,w \in V}
  \sigma_v(x,w)/\sigma(x,w)$. As a slight variation from the classic definition,
  we assume that a node is on its own shortest path:
    $\forall v,w\in V^2,\sigma_v(v,w)/\sigma(v,w)=1$. We adopt the convention:
    $\forall v\in V,\sigma_v(v,v)/\sigma(v,v)=0$. The reason of this variation will become clear
     in the next section.

\item \emph{Closeness Centrality:}
  The \emph{closeness centrality} $\CC$ of a
  node $v\in V$ is defined as: $\CC(G,v)=\sum_{w\in V\setminus v} 2^{-d(v,w)}$.

% \item \emph{Cluster Centrality}:
%   The \emph{cluster centrality} $\KC$ of a
%   node $v\in V$ is defined as the cluster coefficient
%   of $v$, i.e., the number of triangles in which $v$ is involved
%   divided by all possible triangles in $v$'s neighborhood.
\end{enumerate}

By convention, we define the centrality of a node with no edges to be 0.
Moreover, throughout this paper, we will define the graph edit distance between
two graphs $G_1$ and $G_2$ as the minimum number of operations to transform $G_1$ into $G_2$
(or vice versa), where an \emph{operation} is one of the following: link insertion, link removal,
node insertion, node removal.

\section{Graph Distances}\label{sec:graph-dist}

\begin{figure}[t]
\centering
\includegraphics[width=0.35\columnwidth]{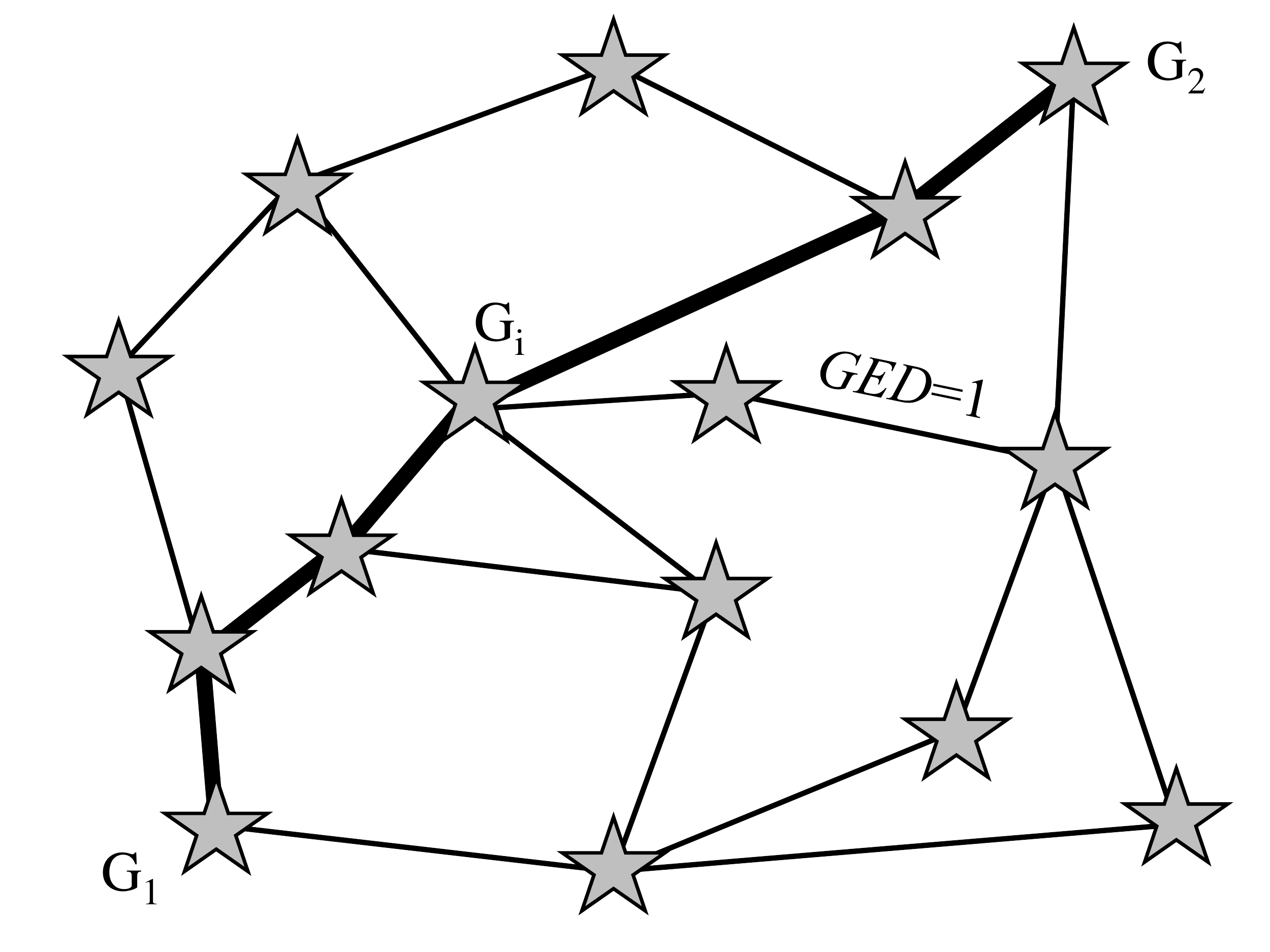}
~~~\includegraphics[width=0.55\columnwidth]{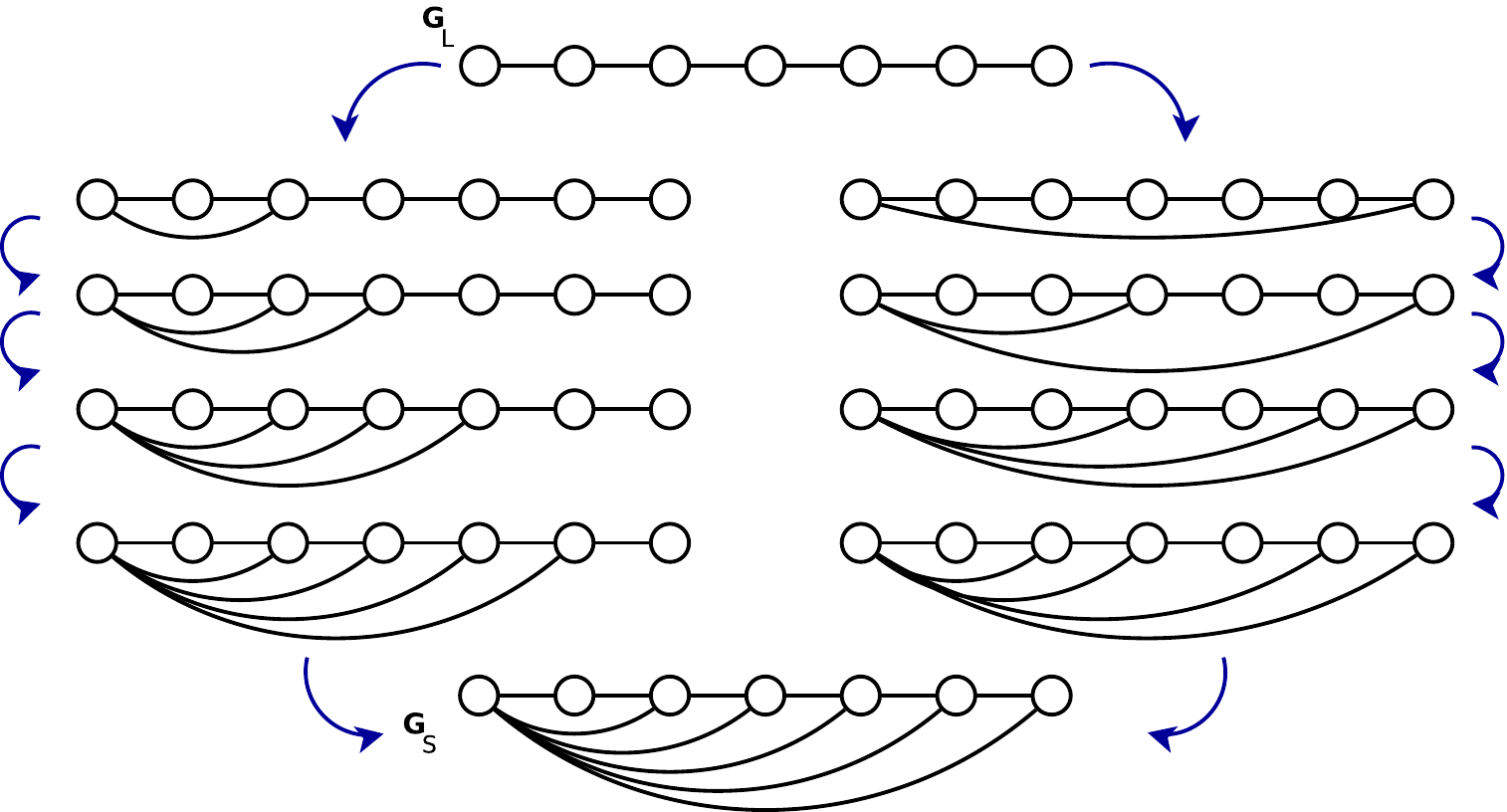}\\
\caption{\emph{Left:} The graph-of-graph $\G$ connects named networks (represented as stars).
Two networks are neighboring iff they differ by a graph edit distance of one.
The centrality distance defines a distance for each pair of neighboring
graphs. \emph{Right:}
Two possible evolutionary paths from a line graph $G_1$ to a shell
  graph $G_2$.}
%\includegraphics[width=0.45\textwidth]{methodology}\\
%Illustration of our methodology: Given an initial graph $G_1$, graphs $G_i$ at a given graph edit distance $d_{\GED}=R$ are
%generated uniformly at random. The actual later graph observed in the experiment is denoted by $G_2$. We test the
%hypothesis that $G_2$ is closer to $G_1$ than other graphs $G_i$ with respect to a certain centrality distance $d_{C}$.}
\label{fig:methodology}
\end{figure}

We now introduce our centrality-based graph similarity measure.
We will refer to the set of all possible topologies by $\G$,
and we will sometimes think of $\G$ being a graph itself:
the ``graph-of-graphs'' which connects graphs with graph edit distance 1.
Figure~\ref{fig:methodology} illustrates the concept.

\begin{definition}[Centrality Distance]\label{def:cd}
   Given a centrality $C$, we define the centrality distance $d_C(G_1,G_2)$
   between two \emph{neighboring} graphs as the component-wise difference:
   $$
   \forall (G_1,G_2) \in E(\G), d_C(G_1,G_2)=\sum_{v\in V} |C(G_1,v)-C(G_2,v)|.
   $$
  % Matthieu I think that's not needed anymore
  % where $C(G_i,v)$ is the centrality of node $v$ in $G_i$, for $i\in\{1,2\}$.

   This definition extends naturally for \emph{non-neighboring} graph couples:
   the distance $d_C(G_1,G_2)$ between $G_1$ and $G_2$ is simply the graph-induced
   distance.
%   $$
%   \forall G_1,G_2 \in \G^2, d_C(G_1,G_2)=\min_{P\in
%     \mathcal{P}_{G_1,G_2}}\sum_{i =1}^{\vert P \vert -1} d_C(P[i],P[i+1]),
%   $$
%   where $\mathcal{P}_{G_1,G_2}$ represents the set of all possible paths
%   between $G_1$ and $G_2$ in $\G$ and $P[i]$ represents the $i^{th}$ graph on
%   path $P$.
 \end{definition}

%\begin{wrapfigure}{r}{0.6\textwidth}
%%\begin{figure}[h]
%\begin{center}
%\vspace{-.7cm}
%\includegraphics[width=0.5\textwidth]{gog}
%	  \caption{The graph-of-graph $\G$ connects named networks (represented as stars).
%Two networks are neighboring iff they differ by a graph edit distance of one.
%The centrality distance defines a distance for each pair of neighboring
%graphs.}\vspace{-.6cm}
%	\label{fig:gog}{fig:gog}{fig:gog}
%	\end{center}
%%\end{figure}
%\end{wrapfigure}

%Informally, the centrality distance between two neighboring graphs (\emph{i.e.}~graphs
% differing just by one edge) is the accumulated change in the
% node centralities.
% The centrality distance between non-neighboring graphs is \emph{induced by the shortest
% distance} along neighboring graphs.

As we will see, the \emph{distance axioms} are indeed fulfilled for the
major centralities. The resulting structure supports the formal study with existing algorithmic tools.
 Let us first define the notion of
 \emph{sensitivity}.

 \begin{definition}[Sensitive Centrality]
   A centrality $C$ is \emph{sensitive} if any single edge modification of any
   graph $G$
   changes the centrality value of at least one node of $G$. Formally, a centrality $C$ is
   \emph{sensitive} iff
   $$\forall G \in \G, \forall e \in E(G), \exists v\in V(G) \textrm{ s.t. } C(G,v) \neq C(G\setminus\{e\},v),$$  where $G \setminus\{e\}$ is the result of removing edge $e$ from $G$.
 \end{definition}

 \begin{lemma}\label{lem:sens}
$\DC$, $\BC$ and $\CC$ are sensitive centralities.
\end{lemma}
%But there are also exceptions, e.g., excentricity.
%: a counter-example graph consists in a lollipop
% graph from which we remove one edge in the fully connected cluster.

It is easy to see that also other centralities, such as cluster centralities
and Page Rank centralities are sensitive.
The distance axioms now follow directly from the graph-induced distance.
 \begin{theorem}\label{thm:dist}
   For any centrality $C$, $d_C$ is a distance on $\G$ iff $C$ is sensitive.
 \end{theorem}

The centrality distance metric of Definition~\ref{def:cd} however comes
with the drawback that it is expensive to compute. Thus, we propose the following approximate version:
\begin{definition}[Approximate Centrality Distance]
   Given a centrality $C$, we define the approximate centrality distance $\widetilde{d_C}(G_1,G_2)$
   between any two graphs as the component-wise difference:
   $$
   \forall (G_1,G_2), \widetilde{d_C}(G_1,G_2)=\sum_{v\in V} |C(G_1,v)-C_2(G_2,v)|.
   $$
%   where $C_i(v)$ is the centrality of node $v$ in $G_i$, for $i\in\{1,2\}$.
\end{definition}

Note that $\widetilde{d_C}\leq d_C$ always holds.
As we will see, while the approximate distance can be far from the exact one
in the worst case, it
features some interesting properties.

% A centrality distance measure $d_C$ could also be used
% to explore the graph-of-graphs, and perform \emph{routing}.
% For example, given two graphs $G_1$ and $G_2$, an evolutionary route can be computed
% as follows: from the current graph $G_t$, the next
% graph $G_{t+1}$ is computed as the neighbor $G_{t+1}\in N(G_t)$ which minimizes
% the centrality distance, i.e., $G_{t+1} = \arg\min_{G'\in N(G_t)} d_C(G',G_t)$.

% \begin{algorithm}[h!]
% \small
% \caption{Greedy Routing: $G_1$ to $G_2$}
%    \label{alg:greedy}
%    \begin{center}
%     \begin{algorithmic}[1]
%         \STATE $G_i:=G_1$;
%         \WHILE {$G_i\neq G_2$}
%             \STATE $G_{i+1} = \arg\min_{G' \in N(G_i)} d(G',G_2)$
%         \ENDWHILE
%     \end{algorithmic}
%     \end{center}
%   \end{algorithm}

% fixme: when to terminate

\section{Example: Closeness Distance}\label{sec:analysis}

To
illustrate the concepts, we will follow the example of a topological evolution
changing a line graph into a shell graph (depicted in
Figure~\ref{fig:methodology} (\emph{right})). This figure shows two different
paths: $A$ incrementally connects node $v_1$ to node $v_i$ for $i=3,\ldots,n$, whereas
path $B$ dichotomically connects $v_i$ to nodes $v_n,v_{n/2},v_{3n/4},v_{n/4},v_{7n/8},v_{5n/8},\ldots$.

\begin{figure}[h]
%\begin{figure}[h]
\begin{center}
%\vspace{-.7cm}
\includegraphics[width=0.5\textwidth]{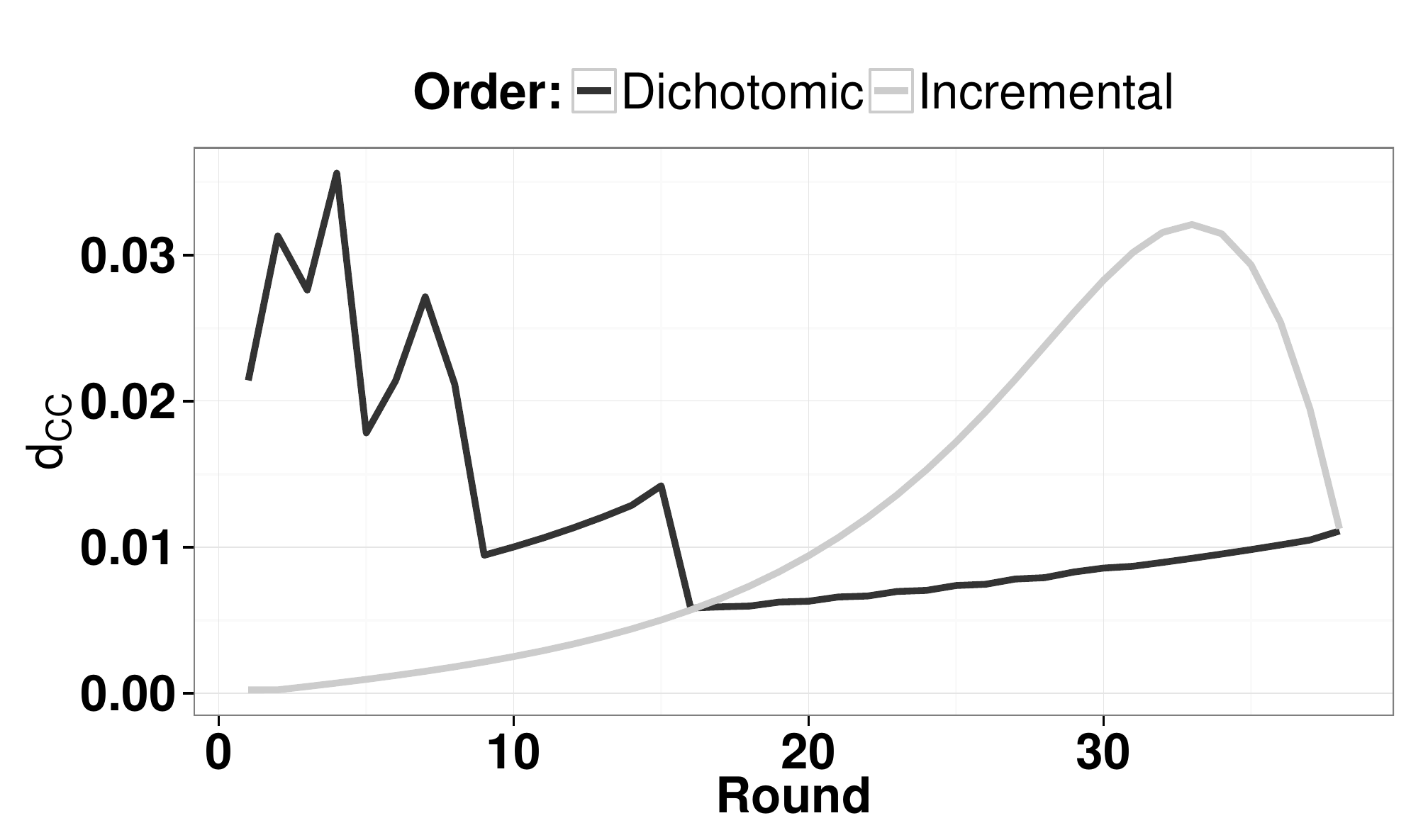}
\caption{Evolution of closeness centrality distance $d_{CC}$ from $G_L$ to
  $G_S$ for $n=40$. Dichotomic and incremental paths are depicted in Figure~\ref{fig:methodology} (\emph{right}).
  Both order define a different path in $\G$ from $G_L$ to $G_S$. At each step, the
  closeness centrality-induced distance between $G_i$ and $G_{i-1}$ is computed.}
\label{fig:ltos2}%
	\end{center}
%\end{figure}
\end{figure}

\subsection{Degree Distance}

As a baseline and for comparison with alternative centralities,
we will consider the degree distance $d_{\DC}(G_1,G_2)$: the distance between two graphs simply counts the number of different edges.
We first make the simple observation that the number of graph edits
is equivalent to the differences in the centrality vectors.
\begin{observation}
The graph edit distance is equivalent to the degree distance, i.e., $d_{\GED} \equiv d_{\DC}$.
\end{observation}

This connection is established by the topological
shortest path in the graph-of-graphs $\G$.
Let us consider our example from Figure~\ref{fig:methodology} (\emph{right}): Since all paths from $G_1$ to $G_2$ which do not introduce
unnecessary edges have the \emph{same cost}, the order in which edges are
inserted is irrelevant. From a $d_\GED$ perspective the incremental (\emph{left}) and
the dichotomic (\emph{right}) paths of Figure~\ref{fig:methodology} (\emph{right}) are
equivalent.
%\gilles{Yet, if you imagine a social network, the steps that would
%  finally make you meet the Pope are likely  incremental.}

We make the following observation:
\begin{observation}
The degree distance resp.~graph edit distance does not provide much insights into
graph evolution paths: essentially all paths have the same costs.
\end{observation}

\subsection{Closeness Distance}

Intuitively, a high closeness distance indicates a large difference in the distances
of the graph.
The closeness distance $d_{\CC}$ has some interesting properties.
For example, the shortest path in the graph-of-graph is connected the the topological shortest path.
In particular, if two graphs are related by inclusion, the shortest closeness path is also a topological
shortest path, as shown in the following.
\begin{theorem}\label{thm:close}
  Let $G_1$ and $G_2$ two graphs in $\G$ such that $E(G_1)\subset E(G_2)$. Then
  all the topological shortest paths (\emph{i.e.} paths that only add edges from
  $G_1$ to $G_2$) are equivalent for closeness.
\end{theorem}

The inclusion property is a relevant property in many temporal networks,
e.g., where links do not
age (\emph{e.g.} if an edge denotes that $u$ has ever met/traveled/read $v$).

To give some intuition of the closeness distance, Figure~\ref{fig:ltos} plots the evolution of the distance terms over time.
As we can see, in the dichotomic order, first longer links are added, while the first incremental steps
are more or less stable.

The dichotomic path has the larger impact on the dynamic graph's shortest path
distances. This indicates that the closeness distance could even be used for ``greedy routing'', in the sense
that efficient topological evolutions can be computed by minimizing the distance to a target topology.
% The late maximum achieved by the incremental order can be understood
%by considering the large amount of neighbors node $v_1$ has at that point: adding
%one edge updates all these distances and therefore has a large impact.

% How will $G_1$ evolve into $G_2$ in our example under closeness
% centrality?
% It turns out that the answer is the same as for betweeness
% centrality.

%\newpage

\section{Experimental Case Studies}\label{sec:experiments}

This section studies the power and limitation of closeness distance empirically, in two case studies: the first scenario is based on
a data set we collected during a cocktail party and models a human mobility pattern;
the second scenario is based on an evolving online social network (OSN) data set which is publicly available.

\textbf{Datasets.}
The first case study is based on the \texttt{SOUK} dataset~\cite{souk}.
This dataset captures the social interactions of 45
individuals during a cocktail, see~\cite{souk} for more details.
The dataset consists in $300$ discrete
timesteps, describing the dynamic interaction graph between the participants, one
timesteps every $3$ seconds.

The second case study is based on a publicly available dataset \texttt{FBL}~\cite{opsahl2009clustering}, capturing all the messages exchanges
realized on an online Facebook-like social network between roughly 20k users
over 7 months. We discretized the data into a dynamic
graph of 187 timesteps representing the daily message exchanges among
users.
For each of these two graphs series, we compare each graph $G_t$ with the
subsequent one: $G_a=G_t$,$G_b=G_{t+1}$. First, we generate a set $S$ of $200$ samples such
that $\forall G_2\in S, d_\GED(G_a,G_2)= d_\GED(G_a,G_b)$. Then we compare the
centrality induced distance $d_C$ from $G_a$ to the samples of $S$ against $d_C(G_a,G_b)$.

\begin{figure}[t]\centering
\begin{center}
%\vspace{-.7cm}
\includegraphics[width=0.405\textwidth]{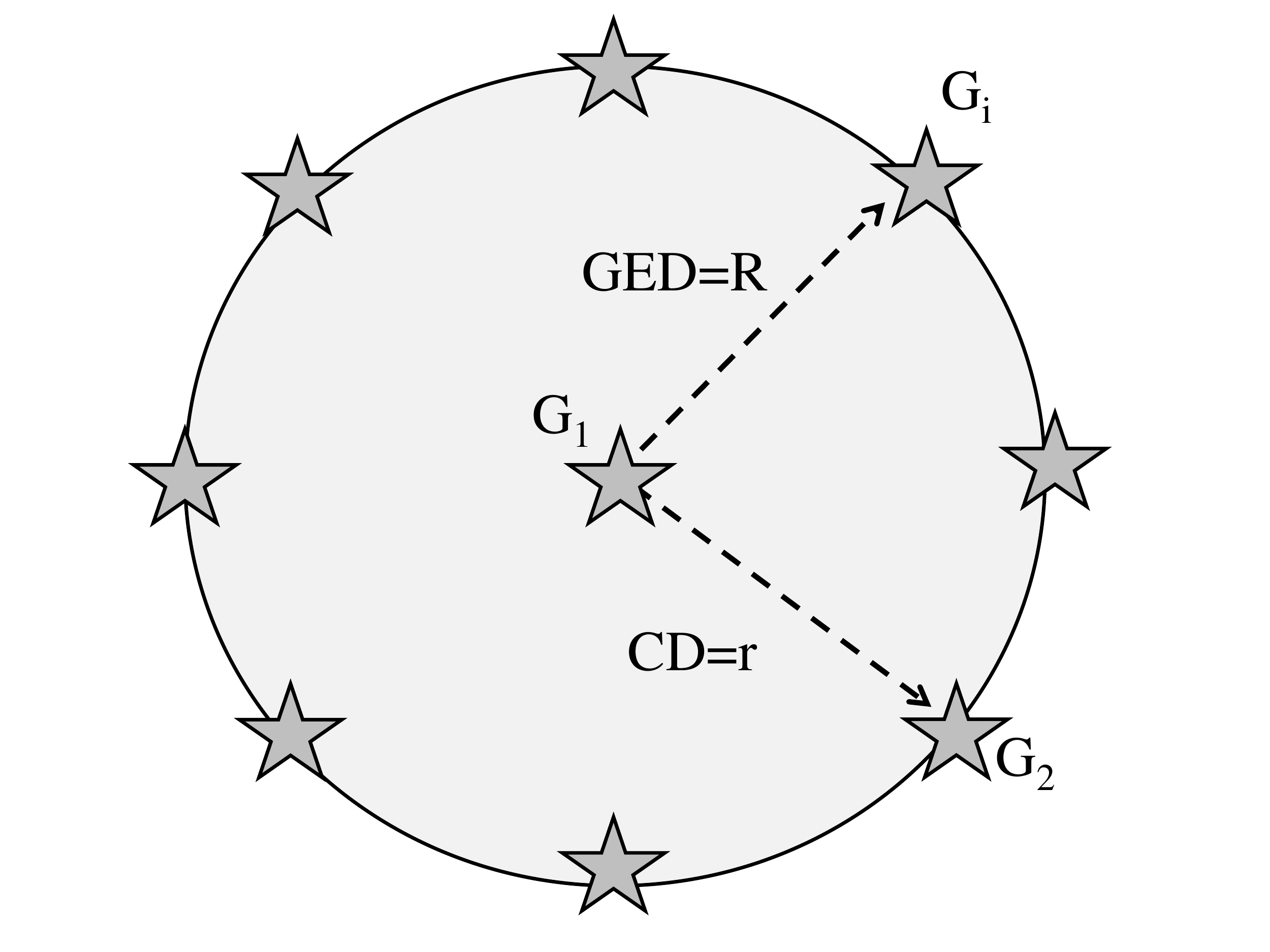}\\
\caption{Illustration of our methodology: Given an initial graph $G_1$, graphs $G_i$ at a given graph edit distance $d_{\GED}=R$ are
generated uniformly at random. The actual later graph observed in the experiment is denoted by $G_2$. We test the
hypothesis that $G_2$ is closer to $G_1$ than other graphs $G_i$ with respect to a certain centrality distance $d_{C}$.}
\label{fig:ltos}
	\end{center}
\end{figure}

\textbf{Methodology.}
We study the question whether centrality distances could be used to predict the
evolution of a temporally evolving network.
To this end, we introduce a simple methodology: We take a graph $G_1$ and a
graph $G_2$ following $G_1$ later in time in the given experiment.
For these two graphs, the graph edit distance (or ``radius'') $R:=d_{\GED}(G_1,G_2)$
is determined, and we generate alternative graphs $G_i$ at the same graph edit
distance $R$ \emph{uniformly at random}.
We investigate the question whether closeness centrality distance can
help to effectively distinguish $G_2$ from other graphs $G_i$,
in the sense that $d_{CC}(G_1,G_2)\ll d_{CC}(G_1,G_i)$ for $i\neq 2$.
Figure~\ref{fig:ltos} illustrates our methodology.
%\begin{wrapfigure}{l}{0.6\textwidth}
%%\begin{figure}[h]
%\begin{center}
%\vspace{-.7cm}
%\includegraphics[width=0.5\textwidth]{methodology}
%	  \caption{Illustration of our methodology: Given an initial graph $G_1$, graphs $G_i$ at a given graph edit distance $d_{\GED}=R$ are
%generated uniformly at random. The actual later graph observed in the experiment is denoted by $G_2$. We test the
%hypothesis that $G_2$ is closer to $G_1$ than other graphs $G_i$ with respect to a certain centrality distance $d_{C}$.}\vspace{-.6cm}
%\label{fig:methodology}
%	\end{center}
%%\end{figure}
%\end{wrapfigure}

\begin{figure*}[t]
\centering
\includegraphics[width=.45\textwidth]{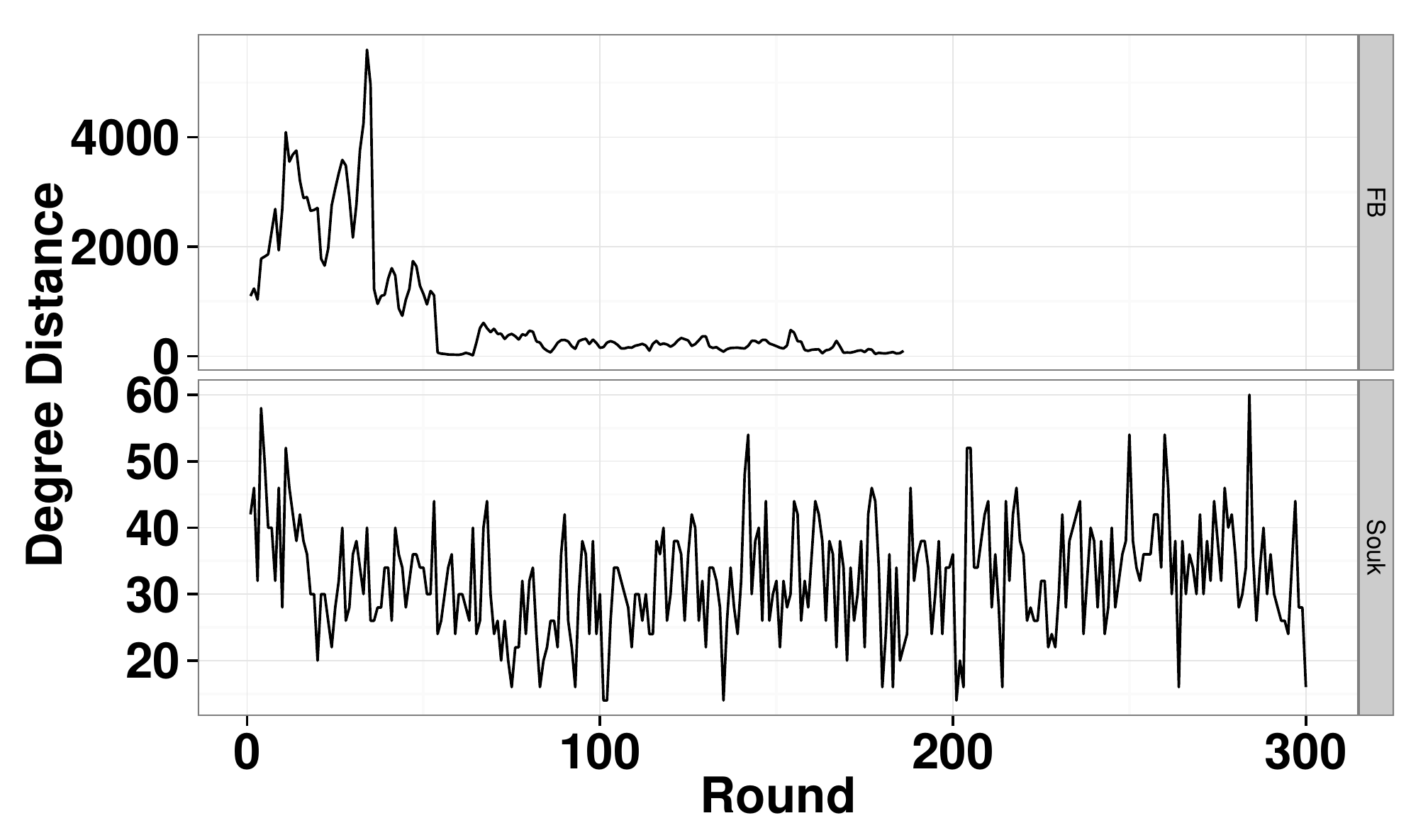}\includegraphics[width=.45\textwidth]{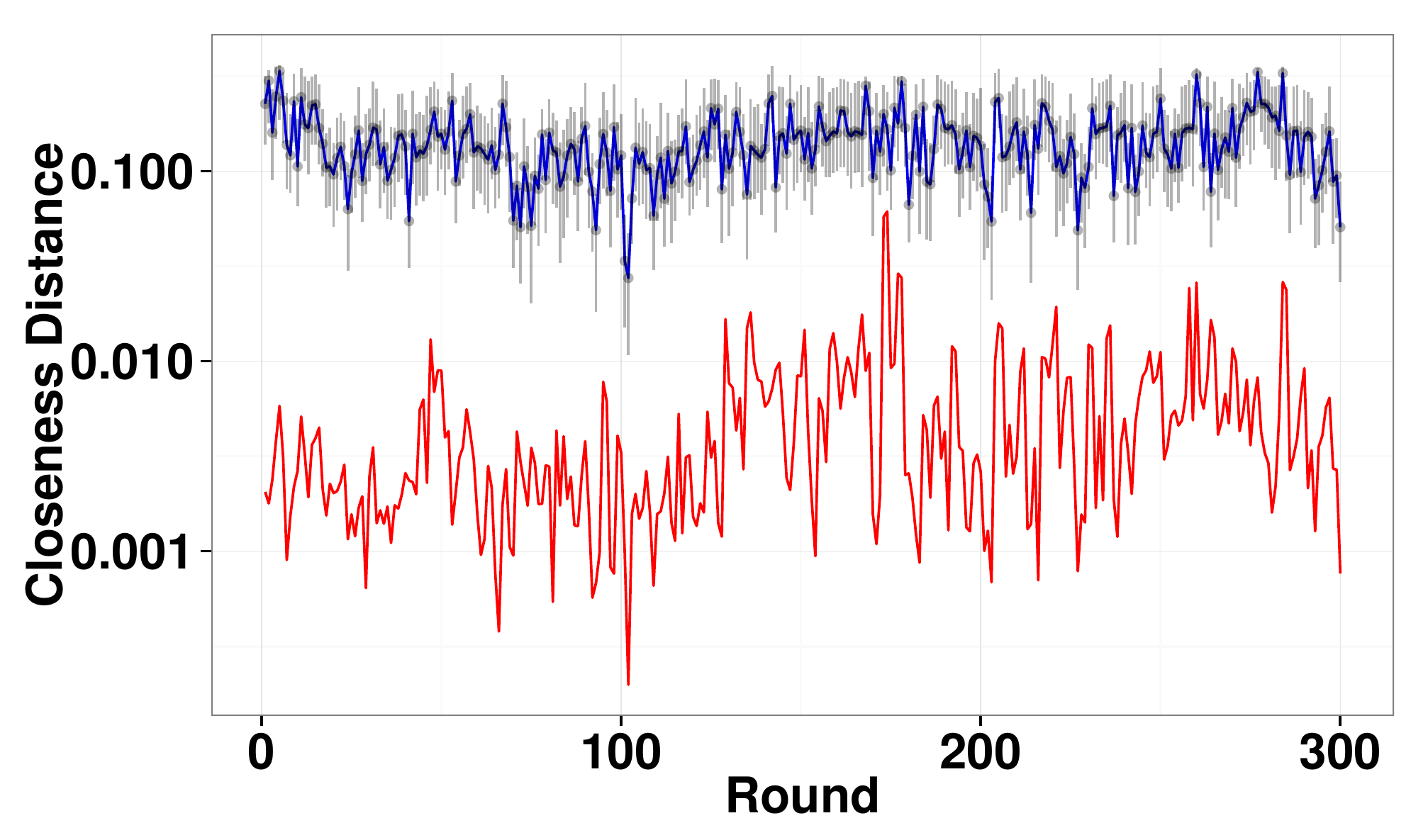}~\\
\caption{\emph{Left:} Measured closeness distance $d_C$ from $G_t$ to $G_{t+1}$
  (\emph{red}), and to $S_{t+1}$ sampled graphs (\emph{blue:} median, \emph{bars:} 5
  and 95 percentiles) over time. \emph{Right:} Temporal representation of the
  measured closeness distance $d_ \CC$ from $G_t$ to $G_{t+1}$ (red) on the
  \texttt{SOUK} dataset, and to $S_{t+1}$ sampled graphs (\emph{blue:} median,
  \emph{bars:} 5 and 95 percentiles) over time.}
\label{fig:temporal}
\end{figure*}
\begin{figure*}[t]
 \centering
\includegraphics[width=0.49\textwidth]{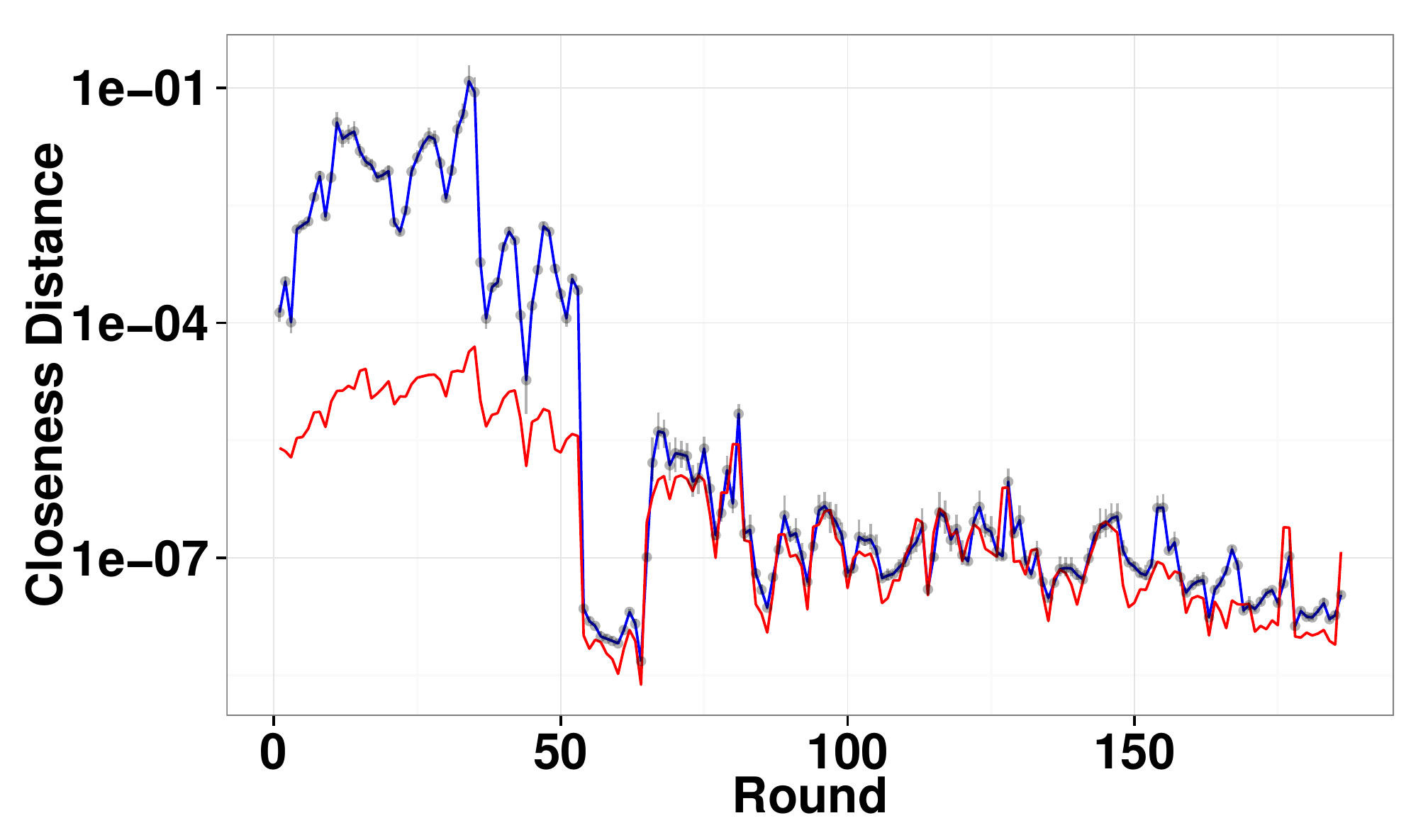}\includegraphics[width=0.49\textwidth]{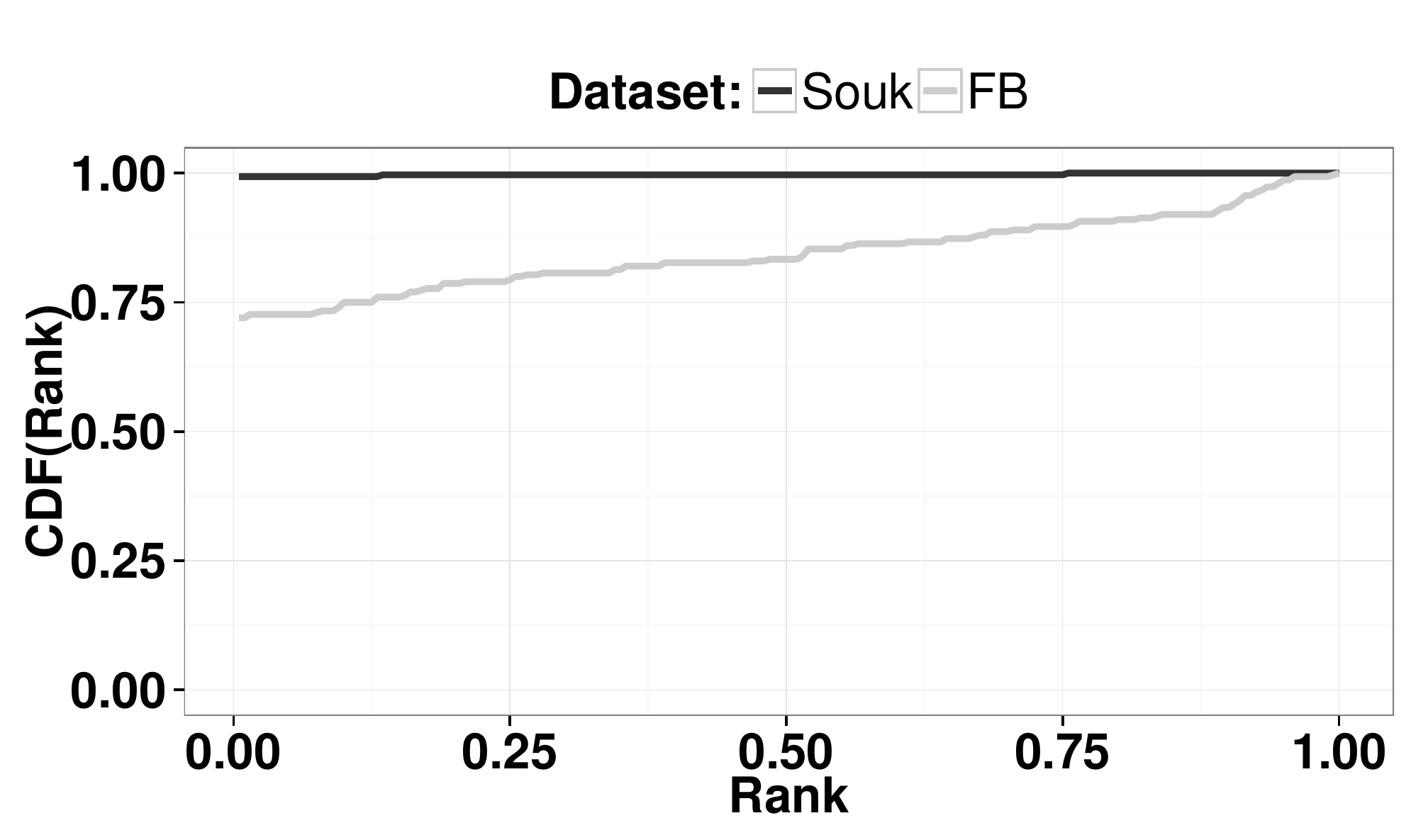}
\caption{\emph{Left:} Temporal representation of the measured closeness distance
  $d_ \CC$ from $G_t$ to $G_{t+1}$ (\emph{red}) on the \texttt{FB} dataset, and to
  $S_{t+1}$ sampled graphs (\emph{blue:} median, \emph{bars:} 5 and 95
  percentiles) over time. \emph{Right:} CDF of the rank of the measured topology
  closeness distance among the closeness distances to the sampled equivalent topologies for both datasets.  }
\label{fig:FB}
\end{figure*}

\textbf{Results.}
Figure~\ref{fig:temporal} (\emph{left}) provides a temporal perspective on the
evolution of $d_\GED$ for both the 300 timesteps of the \texttt{SOUK} dynamic
graph and the 187 timesteps of the \texttt{FB} dataset. Both datasets exhibit
very different dynamics: \texttt{FB} has a high dynamics for the first 50
timestemps, and is then relatively stable, whereas \texttt{SOUK} exhibits a more
regular dynamics.

Figure~\ref{fig:temporal} (\emph{right}) presents the results of our experiment
on the \texttt{SOUK} dataset.  It represents the closeness distance $d_{\CC}$
from each graph $G_t$ to $G_{t+1}$ in red. The distribution of $d_\CC$ values from
$G_t$ to the $200$ randomly sampled graphs of $S_{t+1}$ is represented as
follows: the blue line is the median, while the gray lines represent the $5$ and
$95$ percentiles of the distribution. One can observe that although $\forall G_2
\in S_{t+1}, d_\GED(G_t,G_2)=d_\GED(G_t,G_{t+1})$, most of the time
$d_{\CC}(G_t,G_{t+1})\leq d_{\CC}(G_t,G_2), \forall G_2 \in S_{t+1}$. In other
words, most of the times, the measured graph $G_{t+1}$ is closer to $G_t$ in
closeness distance than the $5\%$ closest randomly sampled graphs.
Figure \ref{fig:FB} (\emph{Left}) presents the same results on the \texttt{FB}
dataset. Here, although most of the time the measured topology is closer in
closeness distance, this is mostly true for the first, most dynamic,
time steps.

Figure~\ref{fig:FB} (\emph{right}) provides a more aggregate view of this
observation: it shows the rank of $d_C(G_t,G_{t+1})$ in the
$\{d_C(G_t,G_2),G_2 \in S_{t+1}\}$ distribution for the closeness ($\CC$)
centrality distance for both dynamic graphs. Values are sorted increasingly, so
rank $0$ represents the smallest. This graph shows that out of the $300$
timesteps of \texttt{SOUK}, the observed graph is closer than any of the 200
random samples in $298$ timesteps for the closeness distance. The same statement holds for
$73\%$ of the $187$ snapshots of the \texttt{FB} dataset.

The observation one can draw from this plot is that when we measure the raw
evolution of topologies (in terms of degree distance) we only grasp a very
incomplete, and rather pessimistic view of the dynamics on both
datasets. Compared to a random evolution that would create the same degree
distance difference, actual topology evolution leaves most of nodes importance
as connectors (closeness distance) unchanged.

\section{Related Work}\label{sec:relwork}

To the best of our knowledge, our paper is the first to combine
the important concepts of graph distances and centralities.
In the following, we will review the related works in the two
fields in turn, and subsequently discuss additional literature on dynamic graphs.

\textbf{Graph distances.}
Graph edit distances have been used extensively
in the context of inexact graph matchings in the
field of pattern analysis.
Central to this
approach is the measurement of the similarity of pairwise
graphs. Graph edit distances are attractive for
their error-tolerance to noise.
We refer the reader to the good
survey by Gao et al.~\cite{ged-survey} for more details.

However, we in this paper argue that the graph edit distance fails to capture
important semantic differences, and are not well suited to measure similarities between complex networks.
Accordingly, we introduce a distance which is based on a parameterizable centrality.

In classic graph theory, notions of similarity often do not take into account
the individual nodes. A special case are graph isomorphism problems:
The graph isomorphism problem is the computational problem of determining whether two finite graphs are isomorphic.
While the problem is of practical importance, and has applications in mathematical chemistry,
many complex networks and especially social networks are inherently non-anonymous. For example, for the
prediction of the topological evolution of a network such as Facebook, or for predicting new topologies based on
human mobility, individual nodes should be taken into account.
 Moreover, fortunately, testing similarity between named graphs is often computationally much more tractable.

\textbf{Graph characterizations and centralities.}
Graph structures are often characterized by the frequency of small patterns
called \emph{motifs}~\cite{motif,motifs,motifs-bioinf,Schreiber05frequencyconcepts}, also known as \emph{graphlets}~\cite{graphlet},
or \emph{structural signatures}~\cite{signatures}.
Another important graph characterization, which is also studied in this paper,
are \emph{centralities}.~\cite{brandes}
Dozens of different centrality indices have been defined over the last years,
and their study is still ongoing, and
a unified theory missing. We believe that our centrality distance
framework can provide new inputs for this discussion.

\textbf{Dynamic graphs.} Researchers have been fascinated by the topological structure and
the mechanisms leading to them for many years. While early works
focused on simple and static networks~\cite{erdos}, later models, e.g.,
based on preferential attachment~\cite{barabasi}, also shed light on
how new nodes join the network, resulting in characteristic graphs.
Nevertheless, today, only very little is known about the dynamics of
social networks. This is also partly due to the lack of good data,
which renders it difficult to come up with good methodologies for
evaluating, e.g., link prediction algorithms~\cite{linkprediction,yang:friendship}.

An interesting related work to ours is by Kunegis~\cite{kunegis}, who also
studied the evolution of networks, but from a spectral graph theory perspective.
In his thesis, he argues that the graph spectrum describes a network on the global level, whereas eigenvectors describe a network at the local level,
and uses these results to devise link prediction algorithms.

\section{Conclusion}\label{sec:conclusion}

We believe that our work opens a rich field
for future research. In this paper, we mainly focused on closeness
distance, and showed that it has interesting properties when applied to the 
use case of dynamic social networks. 
However, our early results indicate that other centralities
have very interesting properties as well.
For instance, it can be seen that using betweenness distance
 to move from a graph $G_1$ to a smaller graph $G_2$ results
 in the same graph sequence as Newman's graph clustering algorithms,
 indicating that betweeness distance can be used to study graph clusterings over time.
The properties, opportunities and limitations of alternative centralities
will be the main focus of our future work.

\textbf{Acknowledgments.}
Stefan Schmid is supported by the DAAD-PHC \emph{PROCOPE} program, the EIT ICT project
\emph{Mobile SDN}, and is part of the INP visiting professor program. Gilles
Tredan is supported by the DAAD-PHC \emph{PROCOPE} program.

%\newpage

%\nocite{*}
{
%\footnotesize
\bibliographystyle{acm}
\bibliography{snds}
\label{sec:References}
}

%\begin{comment}
%\newpage

\begin{appendix}

\section{Proof of Lemma~3.3}

%~\ref{lem:sens}

Let $G \in \G$, and $e=(u,v) \in E(G)$.
\begin{description}
\item {\it Degree centrality:} $\DC(G,u) = \DC(G\setminus\{e\},u) +1 $: \DC{}  is sensitive.
\item {\it Betweenness centrality:} Recall our slightly changed definition
  of betweenness. Now, $\BC(G,u)=\sum_{x,w \in V}
  \sigma_u(x,w)/\sigma(x,w)$. In $G\setminus\{e\}$, all shortest paths are at least as
  long as in $G$, and the shortest path between $u$ and $v$ has increased at
  least one unit: $\BC$ is sensitive.
\item {\it Closeness centrality:}
$\CC(G,u)=\sum_{w\in V\setminus \{u\}} 2^{-d(u,w)}$. In $G\setminus\{e\}$, all distances are greater or equal than in $G$, and strictly greater for the couple $(u,v)$: \CC{} is sensitive.
\end{description}

\section{Proof of Theorem~3.4}

%\ref{thm:dist}

      We show that $d_C$ is a metric on $\G$:
      \begin{description}
      \item[Separation] $\forall G_1,G_2\in \G, d_C(G_1,G_2)\geq 0$ since all summands are non-negative.
      \item[Coincidence] If $G_1=G_2$, we have $\sum_{v\in V}
        |C_1(v)-C_1(v)|=0$. If $G_1\neq G_2$ and $C$ is sensitive, since
        $\forall G \in N(G_1), d_C(G_1,G)>0$, necessarily $d_C(G_1,G_2)>0$.  For the sake of
        contradiction, assume $C$ is not sensitive: $\exists G \in \G, e\in E(G)$
        s.t.~$\forall v \in G, C(G,v)=C(G\setminus\{e\},v)$, and therefore
        $d_C(G,G\setminus\{e\})=0$ with $G\neq G\setminus\{e\}$: $d_C$ is not a metric.
      \item[Symmetry] Straightforward since $|C(G_2,v)-C(G_1,v)|=|C(G_1,v)-C(G_2,v)|$.
      \item[Triangle inequality] Observe that the neighbor-based $d_C$
        definition associates each edge of $\G$ with a strictly positive
        weight. The multi-hop distance $d_C$ is the weighted shortest path in $\G$ given
        those weights. Since the weighted shortest path obeys the triangle
        inequality for strictly positive weights, $d_C$ does as well.
      \end{description}

\section{Proof of Theorem~4.1}

%~\ref{thm:close}

  Let $P$ be a topological shortest paths connecting $G_1$ and
  $G_2$: $ \vert P \vert=d_\GED(G_1,G_2)$. Since $E(G_1)\subset E(G_2)$, $P$
  contains only edge additions to $G_1$, where $\vert P \vert$ denotes
  the path length.
Also $\forall u,v\in V^2, \forall i, d_{P[i]}(u,v)\geq
d_{P[i+1]}(u,v)$, where $P[i]$ denotes the $i$-th node on the path.
Therefore  $\vert  \CC(P[i],(v))-\CC(P[i+1],(v)) \vert =
\CC({P[i]},v)-\CC(P[i+1],v) $.

Therefore:
\begin{align}
  \label{eq:1}
d_{\CC}(G_1,G_2) &\leq& \sum_{i =1}^{\vert P \vert -1} d_{\CC}(P[i],P[i+1])
\nonumber\\
&=&
\sum_{i =1}^{\vert P  \vert -1} \sum_{v\in V}  \CC(P[i],v)-\CC(P[i+1],v)
\nonumber\\
&=&\sum_{v\in V} \CC(G_1,v) - \CC(G_2,v) \nonumber\\ &=&
{\widetilde{d}}_{\CC}(G_1,G_2) \leq d_{\CC}(G_1,G_2) \nonumber\\
\end{align}

  The sum of the closeness distances on this path $P$ is the exact closeness
  distance, for any such path $P$: all shortest paths are equivalent for
  $d_{\CC}$ in this case.
\end{appendix}

%%\nocite{*}
%{\footnotesize \renewcommand{\baselinestretch}{.83}
%\bibliographystyle{plain}
%\bibliography{snds}
%\label{sec:References}
%}
%\end{comment}

\end{document}